\documentclass[letterpaper,twocolumn,10pt]{article}

\usepackage{url}
\usepackage{lmodern,textcomp}
\usepackage{usenix2019_v3}
\usepackage[utf8]{inputenc}

\usepackage{hyperref}
\usepackage{multirow}
\usepackage{tabularx}

\usepackage{booktabs}

\usepackage[underline=false]{pgf-umlsd}
\renewcommand{\mess}[4][0]{
	\path
	(#2)+(0,-\theseqlevel*\unitfactor-0.7*\unitfactor) node (mess from) {};
	\path
	(#4)+(0,-\theseqlevel*\unitfactor-0.7*\unitfactor) node (mess to) {};
	\draw[->,>=angle 60] (mess from) -- (mess to) node[midway, above]
	{#3};
	
	\node (#3 from) at (mess from) {};
	\node (#3 to) at (mess to) {};
}
\newcommand{\messnoarr}[4][0]{
	\path
	(#2)+(0,-\theseqlevel*\unitfactor-0.7*\unitfactor) node (mess from) {};
	\path
	(#4)+(0,-\theseqlevel*\unitfactor-0.7*\unitfactor) node (mess to) {};
	\path[] (mess from) -- (mess to) node[midway, above]
	{#3};
	
	\node (#3 from) at (mess from) {};
	\node (#3 to) at (mess to) {};
}
\newcommand{\messbotharr}[4][0]{
	\path
	(#2)+(0,-\theseqlevel*\unitfactor-0.7*\unitfactor) node (mess from) {};
	\path
	(#4)+(0,-\theseqlevel*\unitfactor-0.7*\unitfactor) node (mess to) {};
	\draw[<->,>=angle 60] (mess from) -- (mess to) node[midway, above]
	{#3};
	
	\node (#3 from) at (mess from) {};
	\node (#3 to) at (mess to) {};
}
\newcommand{\protosep}[1]{
	\begin{pgfinterruptboundingbox}
		\node[text width=.90\textwidth, text height=1pt] at (0.47\textwidth,-\theseqlevel*\unitfactor-1*\unitfactor) {\dotfill #1 \dotfill};
	\end{pgfinterruptboundingbox}
	\stepcounter{seqlevel}
}

\usepackage{bytefield}
\usepackage{xpatch}

\usepackage[n,operators,sets,logic,primitives,keys,adversary]{cryptocode}
\definecolor{darkred}{HTML}{aa0000}
\DeclareMathOperator{\rot}{\pcalgostyle{Rot}}
\DeclareMathOperator{\aenc}{\pcalgostyle{AEnc}}
\DeclareMathOperator{\adec}{\pcalgostyle{ADec}}

\usepackage[style=trad-plain,sortcites,urldate=comp]{biblatex}
\usepackage[acronym,nohypertypes={acronym}]{glossaries}
\makeglossaries
\newacronym{NFC}{NFC}{Near Field Communication}
\newacronym{HCE}{HCE}{Host Card Emulation}
\newacronym{PoC}{PoC}{proof of concept}
\newacronym{PICC}{PICC}{Proximity Integrated Circuit Card}
\newacronym{PCD}{PCD}{Proximity Coupling Device}
\newacronym{APDU}{APDU}{Application Protocol Data Unit}
\newacronym{FWT}{FWT}{Frame Waiting Time}
\newacronym{HAL}{HAL}{hardware abstraction layer}
\newacronym{NCI}{NCI}{NFC Controller Interface}
\newacronym{NFCC}{NFCC}{NFC Controller}
\newacronym{AID}{AID}{Application Identifier}
\newacronym{DB}{DB}{distance bounding}
\newacronym{JNI}{JNI}{Java Native Interface}
\newacronym{ICS}{ICS}{Interindustry Card Standard}

\graphicspath{{media/}}

\def\NFCG{NFCGate}

\begin{document}
	\author{
		{\rm Steffen Klee\footnotemark[1]}\\
		Secure Mobile Networking Lab\\
		TU Darmstadt
		\and
		{\rm Alexandros Roussos\footnotemark[1]}\\
		Secure Mobile Networking Lab\\
		TU Darmstadt
		\and
		{\rm Max Maass}\\
		Secure Mobile Networking Lab\\
		TU Darmstadt
		\and
		{\rm Matthias Hollick}\\
		Secure Mobile Networking Lab\\
		TU Darmstadt
	}

	\title{NFCGate: Opening the Door for NFC Security Research\\with a Smartphone-Based Toolkit}
	\date{}
	
	\maketitle

	\begin{abstract}
	\gls{NFC} is being used in a variety of security-critical applications, from access control to payment systems. However, \gls{NFC} protocol analysis typically requires expensive or conspicuous dedicated hardware, or is severely limited on smartphones. 
	In 2015, the {\NFCG} proof of concept aimed at solving this issue by providing capabilities for NFC analysis employing off-the-shelf Android smartphones.
	
	In this paper, we present an extended and improved \gls{NFC} toolkit based on the functionally limited original open-source codebase.
	With in-flight traffic analysis and modification, relay, and replay features this toolkit turns an off-the-shelf smartphone into a powerful \gls{NFC} research tool.
	To support the development of countermeasures against relay attacks, we investigate the latency incurred by NFCGate in different configurations.
	
	Our newly implemented features and improvements enable the case study of an award-winning, enterprise-level \gls{NFC} lock from a well-known European lock vendor, which would otherwise require dedicated hardware.
	The analysis of the lock reveals several security issues, which were disclosed to the vendor.
	\end{abstract}

	\glsresetall
	
	\renewcommand*{\thefootnote}{\fnsymbol{footnote}}
	\footnotetext[1]{Both authors contributed equally to this research.}
	\renewcommand*{\thefootnote}{\arabic{footnote}}
	
	\section{Introduction}
		
		With the continuous advance of contactless applications, including payment and access control systems, the potential for abuse and security breaches is on the rise.
		This is exemplified by the recent increase of smartphone-based contactless payment transactions using \gls{NFC} \autocite{rl:qz:apay,rl:pay:market} and the expansion of \gls{NFC} capabilities for apps on popular mobile platforms \autocite{rl:apple:nfc,rl:android:nfc}.
		Contactless payment systems have been shown to be susceptible to various attacks \autocite{rl:giese:payment,rl:bocek:payment,rl:jose:payment,rl:roland:payment,rl:roland:payment2,rl:galloway:payment,rl:francis:payment}. 
		Research of potential attacks on \gls{NFC} protocols commonly uses Android devices or dedicated hardware to capture \gls{NFC} traffic. 
		While dedicated hardware provides several advantages in versatility as well as advanced features, such as supported technologies, it is also expensive and can be difficult to use \autocite{chameleon, proxmark}. In contrast, ordinary Android devices look less suspicious in public, are readily available, generally affordable, but usually limited in its set of features compared to dedicated hardware.
		
		{\NFCG}, whose \gls{PoC} was originally presented as a demo by Maass et al. in 2015 \autocite{nfcgate}, is a research toolkit for active \gls{NFC} protocol analysis compatible with the Android platform, which can take advantage of the \gls{NFC} stack and act as a programmable \gls{NFC} device. It provides both the basic features for ordinary Android devices as well as advanced features similar to those found on dedicated hardware, while maintaining availability and affordability. Even though the original \gls{PoC} was able to relay \gls{NFC} traffic, clone card identifiers, and provided basic logging functionality, it had limited compatibility with \gls{NFC} technologies and chipsets.
		
		In this work, we describe the updated version of {\NFCG} \autocite{nfcgate:code}, including its original and new functionality, with regards to networking and \gls{NFC} communication. In particular, our contributions include:
		\begin{itemize}
			\item New modes: replay capabilities for captured \gls{NFC} traffic, on-device mode for capturing traffic system-wide.
			\item Python-based plugin system for analyzing and modifying traffic on-the-fly.
			\item Improvements to the clone mode for duplicating static tag data of different \gls{NFC} technologies.
			\item Extension of the relay mode for wormhole attacks.
		\end{itemize}

		Using {\NFCG}, we also conduct a security analysis of an \gls{NFC}-based smart lock as a case study and evaluate the security of its protocol and implementation to demonstrate the toolkit. Since the lock is not compatible with Android's \gls{NFC} stack by default, this case study would not be possible without {\NFCG}, which enables active attacks on the underlying protocol. Finally, we evaluate the additional latency generally introduced by the use of {\NFCG} in multiple real-world scenarios and discuss potential countermeasures.
	
	\section{Background}
		
		In this section, we provide background information on the techniques and technologies used in {\NFCG}. We discuss \gls{NFC} standards, the \gls{NFC} software stack, and function hooking on Android.
		
		\subsection{NFC Standards}
		The Android \gls{NFC} stack currently supports four basic types of technologies: NFC-A, NFC-B, NFC-F, and NFC-V \autocite{code:nfctech}.
		
		\paragraph{Terminology} While NFC endpoints come in various shapes and sizes, e.g. passive cards, active tags, or even card emulators, the communication in standards is usually defined between two endpoints: \gls{PICC} and \gls{PCD}. A \gls{PICC} is a device in the tag role and a \gls{PCD} is an endpoint in the reader role. While establishing a connection between these endpoints, the \gls{PICC} exchanges some \emph{static tag data} with the \gls{PCD} to initialize the communication. This data depends on the type of \gls{NFC} technology and can be used to help the \gls{PCD} decide between multiple available tags. To exchange data after initialization, messages are sent with the \gls{APDU} encapsulated in their payload.
		
		\paragraph{NFC-A/B} The ISO/IEC~14443 family of standards consists of four parts, describing the different layers in standardized \gls{NFC} communication. While Part~1 describes the physical properties of \gls{NFC} at the lowest level \autocite{std:iso:14443-1}, Part~4 specifies the ``half-duplex block transmission protocol'' \autocite{std:iso:14443} used to transfer \glspl{APDU} between \gls{NFC}-enabled endpoints of type A and B. Several open and proprietary protocols are built on top of this protocol
		and allow for higher-level operations such as authentication, file access, and cryptographic computations to be performed on the tag. 
		For example, one of these open protocols is defined in the ISO/IEC~7816-4 \gls{ICS} \autocite{std:iso:7816-3,std:iso:7816-4},
		while the widely used Mifare DESFire is a proprietary protocol \autocite{std:desfire}.
		
		The transport layer also includes the negotiation of a \gls{FWT}, which is specified by the \gls{PICC} as $i$, and allows the \gls{PCD} to retransmit a message if no response was received within the interval \autocite[Section~7.2]{std:iso:14443}. It is defined as follows:
		\begin{equation*}
			FWT_i = \left(\frac{256 \cdot 16}{13.56\,\mathrm{MHz}}\right) \cdot 2^{i} \qquad 0 \le i \le 14
		\end{equation*}
		
		\paragraph{NFC-F} Similarly, the JIS~X~6319-4 standard specifies, among other things, the physical properties and the transmission protocol for ``high-speed proximity cards'' \autocite{std:jis:6319},
		which Sony FeliCa (NFC-F) complies with. Despite its similarities to ISO/IEC~14443, this standard is incompatible and requires both \gls{PICC} and \gls{PCD} to support the tag technology.
		
		\paragraph{NFC-V} In contrast to the previous technologies, the vicinity standard ISO/IEC~15693 (NFC-V) \autocite{std:iso:15693} achieves significantly more distance than ISO/IEC~14443, but trades some throughput for it. 
		Due to its different physical properties resulting from the operation at a greater distance, this standard is incompatible to ISO/IEC~14443 and does not mandate the implementation of any transport protocol beyond exchanging the static tag data.
		
		\paragraph{NCI} The \gls{NCI} standard \autocite{std:nci} specifies the high-level communication protocol between an NCI-conforming NFC controller and the application processor using various underlying transport mechanisms, e.g., $\mathrm{I}^2\mathrm{C}$.
		Most importantly, NCI defines sending configuration streams to the \gls{NFCC}, which modify a standardized setting with each option in the stream. Setting these options can change the way the NFCC presents itself to other readers during card emulation, including emulating static tag data. While the sending of such streams is not supported as an API by Android, it exists internally and could be accessed through binary instrumentation.
		
		\subsection{Android NFC Stack}
		The Android \gls{NFC} stack in general consists of multiple layers ranging from the underlying \gls{NFC} chipset to the high-level programming interfaces available through the Android SDK.
		
		\begin{figure}
			\def\svgwidth{1\linewidth}
			\centering
			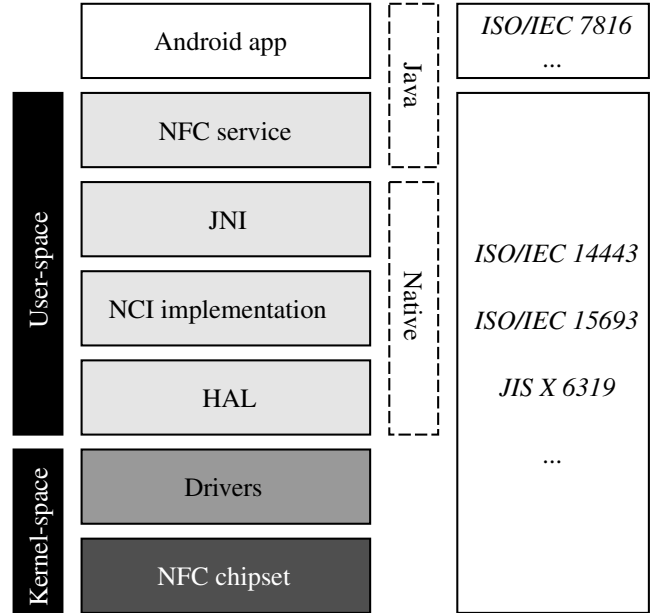
			\caption{Architecture of the Android NFC stack.}
			\label{fig:stack}
		\end{figure}
		
		\autoref{fig:stack} outlines the architecture of the Android \gls{NFC} stack. The lowest level consists of the \gls{NFC} chipset and the \gls{NFC} controller, which interacts with the controller-specific kernel driver, usually via a transport bus like $\mathrm{I}^2\mathrm{C}$ or UART. The communication then leaves the kernel-space and continues in the user-space with the \gls{HAL} that chipset manufacturers are required to implement. This eliminates the need for chipset-specific adjustments of the user-space \gls{NFC} stack. As of now, the Android Open Source Project ships with Broadcom and NXP \gls{HAL} implementations \autocite{code:hal}.
		It consists of native libraries and Java framework classes. Of those native libraries, the \gls{NCI} library communicates with the \gls{HAL} and delivers results to the \gls{NFC} service using the \gls{JNI}. When using the NFC API in the Android SDK, every app communicates with the privileged NFC service.
		
		\subsection{Android HCE}
		\gls{HCE} allows Android-based smartphones with \gls{NFC} to emulate tags for all use-cases where the phone takes the role of the tag, e.g. Google Pay \autocite{rl:android:hce}.
		In order to support multiple \gls{HCE} applications for different purposes on a single smartphone, Android employs a complex \gls{AID} routing mechanism. This requires Android to terminate the ISO/IEC~7816-4 \gls{ICS} connection and intercept the initial \texttt{SELECT} message containing the AID, which is then used to route the requests to the correct application registered for the AID. This artificial limitation prevents Android from accepting any requests not conforming to \gls{ICS} or not starting with the expected \texttt{SELECT} message.
		
		\subsection{Android Function Hooking}
		Hooking is used to intercept regular function calls within foreign applications to modify or extend existing functionality. 
		By injecting instructions into the target function, hooking allows the modification of its control-flow, parameters, return value, and behavior \autocite{rl:lopez:hooking}.
		Because Android applications consist of Java code with optional native code connected via the \gls{JNI}, two types of function hooking exist.
		
		\paragraph{Java Function Hooking.} Since the Java byte code is executed in a Java virtual machine, it is independent of the underlying platform architecture. As a consequence, function hooking in Java is also platform-independent. While the complex process architecture on Android requires significant engineering effort to allow modifications to its virtual machine, the well-known hooking frameworks Xposed \autocite{sw:xposed} and EdXposed \autocite{sw:edxposed} accomplish general Java function hooking by replacing the virtual machine entirely. These frameworks enable application extensions to alter the behavior of any other application on the system.
		
		\paragraph{Native Function Hooking.} In contrast to Java code, native instructions are tied to the processor architecture. Android applications using native code have to provide binaries for all targeted architectures. Consequently, hooking native code requires platform-specific techniques as well. In order to hook native functions, a hooking application substitutes instructions in the target process with the desired behavior. Of the several options that exist for function hooking techniques \autocite{rl:lopez:hooking,rl:vogel:hooking,rl:kim:hooking}, the method we use is \emph{procedure linkage table} hooking provided by the xHook library \autocite{sw:xhook}.
		
	\section{Related Work}

		\begin{table*}
			{\small
			\caption{Feature comparison of {\NFCG} to other NFC tools.}
			\begin{tabularx}{\linewidth}{@{} l l l l l @{}}
				\toprule
				Tool & Protocols & Availability & Usability and Handling & Price\\
				\midrule
				NFCProxy \autocite{nfcproxy}, \autocite{rl:jose:payment,rl:francis:payment} & Only ISO/IEC~7816 APDUs & Android & Inconspicuous, no additional hardware & \$\\
				Proxmark3 \autocite{proxmark} & Any on ISO/IEC~14443 & Dedicated hardware & Suspicious, requires USB host & \$\$\$\\
				ChameleonMini \autocite{chameleon} & Any on ISO/IEC~14443 & Dedicated hardware & Suspicious, requires USB host & \$\$\\
				\textbf{{\NFCG}} & \textbf{Any on ISO/IEC~14443} & \textbf{Android (rooted)} & \textbf{Inconspicuous, no additional hardware} & \textbf{\$}\\
				\bottomrule
			\end{tabularx}
			\label{tab:advantages}
			}
		\end{table*}

		While many software-based tools for the analysis of \gls{NFC}-based protocols have been developed, these tools are mostly tied to their specific use-case. In addition, they operate on high levels in the \gls{NFC} software stack, so that their control of the underlying chipset is limited to that of the available operating system APIs. One of the first software-based \gls{NFC} tools for relaying \glspl{APDU} over the network dates back to 2011 \autocite{rl:francis:payment} and uses a Nokia~6131 as the reader and a BlackBerry~9900 as the card emulator. 
		It shows the applicability of relay attacks on applications using ISO/IEC~14443 \glspl{APDU} and compares timings of smart cards and the relay.
		
		Another software-based tool \autocite{rl:jose:payment} uses \gls{HCE} on the widely available Android operating system to passively (unmodified) relay \glspl{APDU} over the network. With this approach, the tool is able to successfully relay a MasterCard contactless transaction from New York (USA) to Madrid (Spain). Due to limitations on the Android platform, the tool requires the first \gls{APDU} to be a \gls{ICS} \texttt{SELECT} command with a previously registered AID. While a separate Xposed module can be used to circumvent the AID restriction, the tool cannot bypass the \gls{ICS} requirement. Additionally, the tool cannot emulate static tag data due to a lack of platform support.
		
		In a similar way, the discontinued NFCProxy tool \autocite{nfcproxy} uses basic Android \gls{HCE} to relay \gls{NFC} traffic. Due to its short development time, NFCProxy lacks advanced features and bypasses for Android platform restrictions. Despite its shortcomings, NFCProxy is used in a \gls{PoC} attack on the payment systems Visa payWave (qVSDC) and EMV contactless \autocite{rl:bocek:payment}. \autoref{tab:advantages} gives a quick comparison of the aforementioned tools with respect to their features and properties.
		
		Since the introduction of Google Wallet, \gls{NFC}-based mobile payment systems have become a point of interest for security researchers. One particular analysis of Google Wallet \autocite{rl:roland:payment} uses a combination of software- and hardware-based tools to relay \glspl{APDU} from the phone's secure element to a dedicated hardware-based card emulator. This approach requires non-trivial modifications to the operating system in order to access the secure element. Since its publication, Google has fixed this particular attack vector and introduced \gls{HCE}, eliminating the need for using the secure element. Google Wallet (now Google Pay) has since moved to \gls{HCE} \autocite{rl:google:pay}, which has seen widespread usage \autocite{rl:pay:market}.
		
		With the rise of electronic vehicles in Germany, a surge of integrated and cheap charging stations from different providers attracts the interest of security researchers. Due to the simplicity of the charging stations, they do not feature regular payment processing. Instead, many providers rely on lightweight customized billing systems with a supplied token for user authentication. For example, some providers use \gls{NFC}-based smartcards as hardware tokens as analyzed by Dalheimer \autocite{dalheimer:eauto}. He shows that despite the usage of Mifare Classic tags supporting cryptographic operations, some systems only use static tag data as authentication for billing. This allows cloning the tag using a hardware-based \gls{NFC} tool \autocite{chameleon,proxmark}, which emulates the static tag data.
		
		The security of contactless payment systems generally relies on the assumption of the limited range of \gls{NFC}. However, the ReCoil attack \autocite{rl:yuyi:range} shows that this premise does not hold. Using a passive (unpowered) relay consisting of antenna coils and a waist band, the \gls{NFC} communication range can be extended up to $49.6\,\mathrm{cm}$. This allows an attacker to relay the \gls{NFC} traffic over a distance, thus bypassing physical security measures.		
	
	\section{Implementation}
		While the Android operating system acts as an integrated \gls{NFC} reader enabling apps to communicate with tags freely, tag emulation is much more limited. The \gls{HCE} functionality of Android allows the emulation of applications based on ISO/IEC~7816 \gls{ICS}, but has no option to mimic static tag data, which restricts its functionality for \gls{NFC} security analysis.
		{\NFCG} is an Android application that circumvents these restrictions by using symbol hooking in managed (Java) and native (C/C++) code to gain low-level access to the Android \gls{NFC} stack. It is compatible with devices supporting the Xposed \autocite{sw:xposed} or EdXposed \autocite{sw:edxposed} hooking framework.
		
		As the code of {\NFCG} has undergone many changes, we focus on the technical implementation of the following features of {\NFCG}, where \autoref{tab:compare} distinguishes the original \gls{PoC} from the current version of {\NFCG}:
		
		\begin{itemize}
			\item \emph{Standardized logging format}. Logging \gls{NFC} traffic in the standardized packet capture format \emph{pcapng} \autocite{spec:pcapng} enables the use of packet analyzers with advanced dissectors, such as Wireshark \autocite{sw:wireshark}.
			\item \emph{Clone mode.} The clone mode allows the emulation of static tag data for NFC-A, NFC-B, and NFC-F technologies.
			\item \emph{Relay mode.} The relay mode features a server software with a Python plugin system for analyzing and modifying traffic on-the-fly.
			\item \emph{Replay mode.} This mode replays previously recorded or imported traffic in either reader or tag role locally without a relay setup, or remotely with on-the-fly traffic modifications.
			\item \emph{On-device capture mode.} Capturing system-wide \gls{NFC} traffic from other apps running on the Android device allows the analysis of \gls{HCE}-based applications as well as \gls{NFC} reader applications.
		\end{itemize}
	
		\begin{table}[b]
		{\small
			\caption{Comparing original \gls{PoC} to current {\NFCG} version.}
			\begin{tabularx}{\linewidth}{@{} l l l @{}}
				\toprule
				 & \gls{PoC} & Current version of {\NFCG}\\
				\midrule
				OS version & max. 6 & max. 10\\
				Architecture & ARMv7 & Added ARM64\\
				Chipsets & Broadcom & Any \gls{NCI}\\
				Technologies & A & Added B, F\\
				Modes & Clone, relay & Added replay, on-device capture\\
				Interoperability & - & \multirow{2}{4cm}{Added logging, import/export, Python plugin system}\\
				& & \\
				\bottomrule
			\end{tabularx}
			\label{tab:compare}
		}
		\end{table}
	
	\subsection{Logging and Interoperability}\label{sec:pcapng}
		When {\NFCG} processes \gls{NFC} traffic, it is logged to the app's private storage in a database. The app provides an overview of these logs, their timestamps, and the mode they have been recorded in. A single log shows \glspl{APDU} of the \gls{NFC} traffic with their timestamp, a hexadecimal binary dump of data, and an indicator of the originator device (reader or tag).
		
		{\NFCG} allows exporting and importing logs in the \emph{pcapng} file format, providing interoperability with other tools such as the protocol analyzer Wireshark \autocite{sw:wireshark}. In order to encode captured \glspl{APDU}, ISO/IEC~14443 framing is used with the predefined \texttt{DLT\_ISO\_14443} link type. Since no predefined link type exists for storing static tag data, the user-defined link type \texttt{DLT\_USER\_0} is used instead. Because \glspl{APDU} are captured without their associated header or checksum, {\NFCG} creates an artificial header of type \texttt{I\_BLOCK} with a direction indicator of either \gls{PICC}$\rightarrow$\gls{PCD} or \gls{PCD}$\rightarrow$\gls{PICC} and a flag bit to ignore the missing checksum. This framing of \glspl{APDU} with ISO/IEC~14443 headers enables the use of Wireshark's protocol dissector \autocite{ws:iso14443} and its sub-dissectors such as ISO/IEC~7816 \autocite{ws:iso7816} for analysis.
	
	\subsection{Clone Mode}\label{sec:clone}
		Despite the \gls{HCE} API on Android not supporting static tag data emulation, the lower-level NCI stack allows setting arbitrary tag data. Using symbol hooking, {\NFCG} gains access to the lower-level NCI stack, which allows the \emph{clone mode} to emulate any captured static tag data. Note that while even some dedicated hardware restricts the range of values allowed to be manually set, no such restrictions were encountered using the Android NCI stack. 
				
		The NCI standard defines the \texttt{CORE\_SET\_CONFIG\_CMD} command to configure the \gls{NFC} discovery in poll and listen mode. When emulating a tag (listen mode), the NFCC uses the supplied configuration stream to initialize itself. In the NCI implementation \autocite{code:native}, the \texttt{NFC\_SetConfig} function is used for sending a configuration stream to the NFCC. Using this function with custom configuration options allows {\NFCG} to set the static tag data.
		
		\begin{figure}
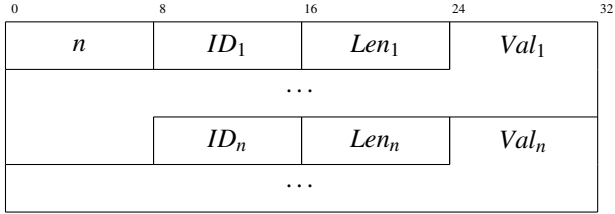

			\centering
			{
				\begin{bytefield}[bitwidth=.7em]{32}
					\bitheader{0,8,16,24,32}\\
					\bitbox{8}{$n$}
					\bitbox{8}{$ID_1$} & \bitbox{8}{$Len_1$} & \bitbox[ltr]{8}{$Val_1$}\\
					\wordbox[lr]{1}{\ldots}\\
					\bitbox[lbr]{8}{}\bitbox{8}{$ID_n$} & \bitbox{8}{$Len_n$} & \bitbox[ltr]{8}{$Val_n$}\\
					\wordbox[lrb]{1}{\ldots}
				\end{bytefield}
			}
			\caption{Format of NCI configuration parameters.}
			\label{fig:nci:config}
		\end{figure}
	
		\begin{table}
			\caption{Available NCI config parameters per technology.}
			\begin{tabularx}{\linewidth}{@{} l l >{\raggedright\arraybackslash}X @{}}
				\toprule
				Tech. & Prefix & Parameter $IDs$\\
				\midrule
				NFC-A & \texttt{LA\_} & \texttt{NFCID1}, \texttt{SEL\_INFO}, \texttt{BIT\_FRAME\_SDD}, \texttt{PLATFORM\_CONFIG}, \texttt{HIST\_BY}\\
				NFC-B & \texttt{LB\_} & \texttt{NFCID0}, \texttt{APPLICATION\_DATA}, \texttt{SFGI}, \texttt{SENSB\_INFO}, \texttt{ADC\_FO}, \texttt{H\_INFO\_RSP}\\
				NFC-F & \texttt{LF\_} & \texttt{T3T\_IDENTIFIERS\_1}, \texttt{T3T\_FLAGS}, \texttt{T3T\_PMM}\\
				\bottomrule
			\end{tabularx}
			\label{tab:nci:config}
		\end{table}
		
		\autoref{fig:nci:config} shows the format of the configuration stream. $n$ is the total number of options in the stream, while $ID_i$, $Len_i$, and $Val_i$ specify the identifier, length in octets, and value of the configuration parameter. \autoref{tab:nci:config} lists the relevant parameter $IDs$ for the emulation of static tag data in any technology supported by Android \gls{HCE}.
		
		The NCI library also invokes \texttt{NFC\_SetConfig} at different times, such as when starting RF discovery, overwriting the configuration set by {\NFCG}. In order to protect its custom set configuration and increase reliability, this function is altered using symbol hooking. The hook removes configuration parameters that would overwrite the custom values before passing them to the original function. This not only ensures that the custom configuration stays active at all times, but also allows {\NFCG} to save the rejected values and apply them after closing \emph{clone mode} to restore the NFCC to its normal operating state.
		
		While this mode allows the emulation of static tag data, it does not support responding to any \gls{APDU} commands. Despite this restriction, \emph{clone mode} is useful when static tag data (i.e., the NFCID) is used for access control as seen in \autocite{dalheimer:eauto}, since it only requires one device running {\NFCG}.
	
	\subsection{Relay Mode}
		
		\begin{figure}
			\def\svgwidth{\linewidth}
			\centering
\begingroup%
  \makeatletter%
  \providecommand\color[2][]{%
    \errmessage{(Inkscape) Color is used for the text in Inkscape, but the package 'color.sty' is not loaded}%
    \renewcommand\color[2][]{}%
  }%
  \providecommand\transparent[1]{%
    \errmessage{(Inkscape) Transparency is used (non-zero) for the text in Inkscape, but the package 'transparent.sty' is not loaded}%
    \renewcommand\transparent[1]{}%
  }%
  \providecommand\rotatebox[2]{#2}%
  \newcommand*\fsize{\dimexpr\f@size pt\relax}%
  \newcommand*\lineheight[1]{\fontsize{\fsize}{#1\fsize}\selectfont}%
  \ifx\svgwidth\undefined%
    \setlength{\unitlength}{261bp}%
    \ifx\svgscale\undefined%
      \relax%
    \else%
      \setlength{\unitlength}{\unitlength * \real{\svgscale}}%
    \fi%
  \else%
    \setlength{\unitlength}{\svgwidth}%
  \fi%
  \global\let\svgwidth\undefined%
  \global\let\svgscale\undefined%
  \makeatother%
  \begin{picture}(1,0.36206897)%
    \lineheight{1}%
    \setlength\tabcolsep{0pt}%
    \put(0,0){\includegraphics[width=\unitlength,page=1]{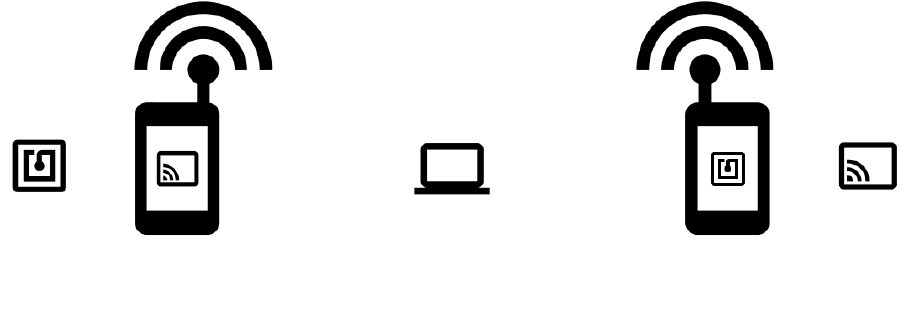}}%
    \put(0.01804922,0.03312841){\makebox(0,0)[lt]{\lineheight{1.25}\smash{\begin{tabular}[t]{l}(A)\end{tabular}}}}%
    \put(0.17012094,0.03312841){\makebox(0,0)[lt]{\lineheight{1.25}\smash{\begin{tabular}[t]{l}(B)\end{tabular}}}}%
    \put(0.77607902,0.03312841){\makebox(0,0)[lt]{\lineheight{1.25}\smash{\begin{tabular}[t]{l}(D)\end{tabular}}}}%
    \put(0.93245022,0.03312841){\makebox(0,0)[lt]{\lineheight{1.25}\smash{\begin{tabular}[t]{l}(E)\end{tabular}}}}%
    \put(0.47342901,0.03312841){\makebox(0,0)[lt]{\lineheight{1.25}\smash{\begin{tabular}[t]{l}(C)\end{tabular}}}}%
    \put(0,0){\includegraphics[width=\unitlength,page=2]{relay.pdf}}%
  \end{picture}%
\endgroup%

			\caption{NFCGate relay setup with external server.}
			\label{fig:relay}
		\end{figure}
	
		Extending clone mode, the \emph{relay mode} forwards received \glspl{APDU} and static tag data to a server, as well as sending \glspl{APDU} received from the server back to the tag or reader. \autoref{fig:relay} shows the interaction of a legitimate tag (A), a legitimate reader (E), and two devices running {\NFCG} in reader (B) and tag mode (D) connected to the server (C). When the server receives data from {\NFCG}, it now processes the data by filtering it through all enabled plugins before broadcasting the potentially modified data to all other connected {\NFCG} devices in the same session. Plugins are invoked in a user-specified order and able to fully modify data, in accordance with the Dolev-Yao model \autocite{dolevyao}.
		The server is written in the Python programming language and does not require external dependencies, which enables hosting the server directly on smartphones with any Python interpreter.
		
		In relay mode, {\NFCG} devices choose between the two roles \emph{reader} and \emph{tag}, while logging all data in the \emph{pcapng} format (see \autoref{sec:pcapng}). Using two connected {\NFCG} devices, one as the reader and one as the tag, in a setup similar to \autoref{fig:relay}, enables a user to capture \gls{NFC} traffic between a legitimate tag and reader in NFC-enabled systems.
			
	\subsection{Replay Mode}
		In \emph{replay mode}, {\NFCG} uses previously captured logs to communicate with an \gls{NFC} device. As in relay mode, it supports the two roles \emph{reader} and \emph{tag}. The role specifies which side of the recorded \gls{NFC} traffic to replay to the device. For example, when using the tag role, {\NFCG} emulates a tag, similar to clone mode. When the reader sends \glspl{APDU} to {\NFCG}, it responds with the corresponding tag \gls{APDU} taken from the selected log. Multiple options exist for the selection of the \gls{APDU} to replay: In \emph{index-based} mode, the selection is only based on the position in the log, while the \emph{data-based} replay selects responses based on the contents of the request \gls{APDU}. All replay traffic is recorded in a new log for further analysis. In addition to replaying unmodified traffic from the log, {\NFCG} supports \emph{advanced replay} over the network. This setting routes all traffic over the server, allowing for modifications or replacements of the replayed traffic by server plugins before sending it to the \gls{NFC} device. By not requiring a second device running {\NFCG}, active attacks on an \gls{NFC} protocol using replay mode have a lower response latency than attacks in relay mode, which could potentially bypass relay countermeasures based on timing.

	\subsection{On-Device Capture}
		The \emph{on-device capture mode} supports capturing low-level \gls{NFC} traffic of the Android device while reading or emulating an \gls{NFC} tag. {\NFCG} hooks central functions in the \gls{NFC} service, allowing it to capture the entire traffic without interfering with other apps. This enables undetected capturing of the \gls{NFC} traffic from other apps without requiring any relay setup, thus avoiding problems with latencies introduced by a relay.
		
		We hook the \texttt{transceive} method of the \texttt{NFCService} \autocite{code:transceive} after its execution to capture \gls{NFC} traffic of the device while reading a tag. This method receives data to be sent to the currently active tag, with the tag's response in the return value. When the device discovers a tag, it invokes the \texttt{dispatchTag} method of \texttt{NfcDispatcher} \autocite{code:dispatch}, which is hooked to enable capturing the static tag data.
		
		In \gls{HCE} mode, the \texttt{HostEmulationManager} \autocite{code:hce} manages the routing of data to apps registered with different \glspl{AID}. On discovery of the tag device by an external reader, \texttt{onHostEmulationActivated} is called. Whenever the reader transmits data, \texttt{onHostEmulation\-Data} receives this data and routes it to the correct application. Hooking this method captures \gls{NFC} traffic from a reader to the device in \gls{HCE} mode, even if no \gls{AID} is registered or the \gls{APDU} is not compatible with ISO/IEC~7816-4. This circumvents the Android system restriction on the initial reader \gls{APDU}.
		Any response of the \gls{HCE} application is routed through the \texttt{NFCService.sendData} function, hence hooking this function also enables capturing \gls{NFC} traffic from the device to the reader.
		
		All hooks collect the captured data locally within the context of the \gls{NFC} service. The {\NFCG} app requests these captures from the service through broadcasts and intents. In order to be exported and used in any other {\NFCG} mode, the capture is logged in the context of the app.

\section{Case Study: Smart Door Lock}%
In order to demonstrate the capabilities of {\NFCG} in analyzing \gls{NFC} protocols using off-the-shelf smartphones, we conduct a case study on an \gls{NFC}-based locking system. 

\subsection{Overview}
The subject of our case study is an enterprise-level electronic access control system manufactured by a well-known European security vendor\footnote{Results published with consent of the vendor.}.
We choose this product based on the existing reputation of the vendor for physical locking mechanisms, the security award the product had received, and the fact that it was available as an off-the-shelf solution unlike, for example, NFC-based hotel door locking systems.
The system consists of a base station and cylinder locks, with permissions for each component bound to user accounts manageable via a web interface. A user issues actions such as unlocking doors with a transponder or an app compatible with the Android and iOS operating systems. In this paper, we focus on the NFC-based cylinder unlocking mechanisms, leaving aside communication with the base station. Our hardware setup consists of the base station, one cylinder lock, and two DESFire transponders. In addition to the hardware, we use the reader software, which is capable of adding transponders to the deployment.

The base station contains the web-based user, cylinder, and permission management. The lock connects to the base station wirelessly with a proprietary protocol in the 868.3 MHz band, requesting authorization and configuration parameters. The transponder presents itself to the lock using the \gls{NFC} protocol described in \autoref{sec:lock} when in range. Then, the lock checks the unique identifier $UID$ obtained from the transponder by sending this information to the base station. If this $UID$ is authorized to access the lock, it accepts the transponder and unlocks the cylinder.

	\subsection{Unlocking Procedure}\label{sec:lock}
	Using the relay mode of {\NFCG}, several communications between lock and transponder were recorded and exported in the \emph{pcapng} format. Even though the locking protocol does not conform to \gls{ICS}, {\NFCG} circumvents the Android limitation and allows us to receive arbitrary \glspl{APDU}. Analyzing the captured traffic using a combination of Wireshark and the libfreefare source code~\autocite{sw:freefare} as a reference, we were able to reverse-engineer the protocol used by the locking system.

	The lock uses a modified DESFire AES authentication protocol to derive a session key and to establish an authenticated and encrypted channel to the transponder. It is not immediately clear why the protocol was modified. Using the established channel, the transponder transmits its $UID$ to the lock securely.
	
	\begin{figure}[b]
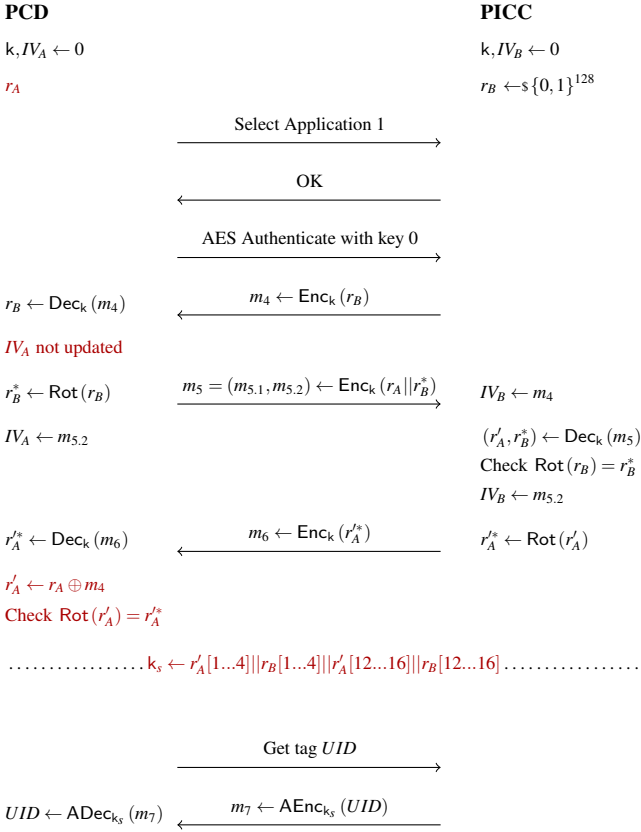

		\centering
		\procedure[addtolength=5pt,width=\columnwidth,codesize=\scriptsize]{}{%
			\footnotesize\textbf{PCD} \< \< \footnotesize\textbf{PICC}\\
			\key, IV_A \gets 0 \< \< \key, IV_B \gets 0\\
			\textcolor{darkred}{r_A}\< \< r_B \sample \bin^{128}\\
			\< \sendmessageright*{\text{Select Application 1}} \<\\
			\< \sendmessageleft*{\text{OK}} \<\\
			\< \sendmessageright*{\text{AES Authenticate with key 0}}\<\\
			r_B \gets \dec_{\key}\left(m_4\right) \< \sendmessageleft*{m_4 \gets \enc_{\key}\left(r_B\right)}\<\\
			\text{\textcolor{darkred}{$IV_A$ not updated}}\<\<\\
			r_B^* \gets \rot{\left(r_B\right)} \< \sendmessageright*{m_5 = \left(m_{5.1}, m_{5.2}\right) \gets \enc_{\key}\left(r_A || r_B^*\right)} \< IV_B \gets m_4\\
			IV_A \gets m_{5.2}\< \< \left(r_A',r_B^*\right) \gets \dec_{\key}\left(m_5\right)\\
			\< \< \text{Check } \rot{\left(r_B\right) = r_B^*}\\
			\< \< IV_B \gets m_{5.2}\\
			r_A'^* \gets \dec_{\key}\left(m_{6}\right)\< \sendmessageleft*{m_6 \gets \enc_{\key}\left(r_A'^*\right)} \< r_A'^* \gets \rot{\left(r_A'\right)}\\
			\textcolor{darkred}{r_A' \gets r_A \xor m_4} \<\<\\
			\textcolor{darkred}{\text{Check } \rot{\left(r_A'\right) = r_A'^*}} \<\<\pclb
			\pcintertext[dotted]{\textcolor{darkred}{$\key_s \gets r_A'[1...4] || r_B[1...4] || r_A'[12...16] || r_B[12...16]$}}\\
			\< \sendmessageright*{\text{Get tag $UID$}} \< \\
			UID \gets \adec_{\key_s}\left(m_7\right) \< \sendmessageleft*{m_7 \gets \aenc_{\key_s}\left(UID\right)} \< \\
		}
	\caption{Cylinder unlocking procedure.}%
		\label{proto:lock}
	\end{figure}

	\autoref{proto:lock} shows the protocol for the unlocking procedure. Deviations from the DESFire AES authentication protocol are marked in red. The key $k$ is a pre-shared 128-bit AES key. The encryption and decryption routines, $\enc$ and $\dec$, use AES-128 in CBC mode. The initialization vector (IV) for the cryptographic operations is stored in the variables $IV_A$ and $IV_B$. The secure channel uses AES-128 in CBC mode with CMAC for authentication in the routines $\aenc$ and $\adec$.
	
	Initially, the \gls{PICC} picks a random nonce $r_B$ and sends it to the \gls{PCD} encrypted with the pre-shared key $k$. Receiving this message, \gls{PCD} decrypts $r_B$ and sends its own nonce $r_A$ and the rotated $r_B^*$ to \gls{PICC}. At this point, the protocol deviates from DESFire authentication by using a static value for $r_A$ and not updating the IV used for encryption. This results in message $m_{5.1}$ being encrypted under $IV = 0$ while \gls{PICC} expects $IV = m_4$. Therefore, \gls{PICC} decrypts $m_{5.1}$ with $IV = m_4$ and obtains $r_A' = r_A \xor m_4$ with $r_A' \ne r_A$. If the decrypted $r_B^*$ matches the local rotation of $r_B$ the \gls{PICC} accepts the authentication and derives the session key $k_s$. Finally, it sends the rotated $r_A'^*$ to the \gls{PCD}. In order to reconcile the difference between $r_A'^*$ and the expected $r_A^*$, the \gls{PCD} needs to calculate $r_A' = r_A \xor m_4$. After checking that the received $r_A'^*$ matches the local rotation of $r_A'$, the \gls{PCD} accepts the authentication and derives the session key $k_s$. Since the \gls{PICC} only knows $r_A'$, the session key must be derived from it instead of $r_A$.
	
	Concluding the interaction, the \gls{PICC} sends its $UID$ over the encrypted and authenticated channel using the established session key $k_s$. All encrypted channel communication contains a CMAC-based authentication tag alongside the plaintext, in accordance with the DESFire AES authentication protocol.

	\begin{figure*}
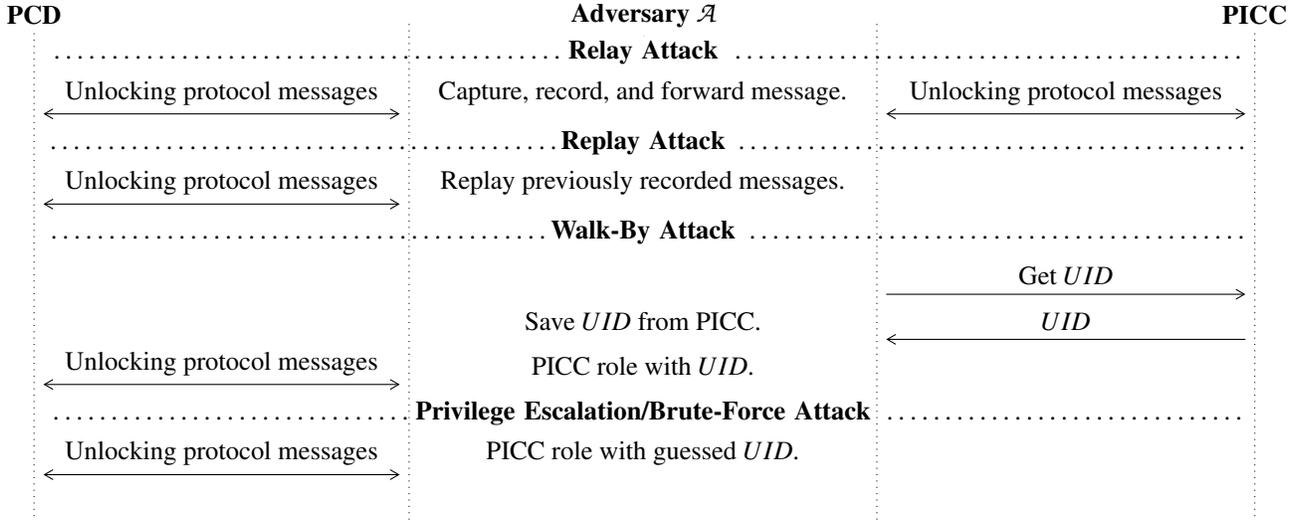

		\centering
		\begin{sequencediagram}
			\tikzstyle{inststyle} = []
			\newinst{pcd}{\textbf{PCD}}
			\newinst[4.5]{adv}{}
			\newinst[3.0]{advm}{\textbf{Adversary \adv}}
			\newinst[2.0]{adve}{}
			\newinst[4.9]{picc}{\textbf{PICC}}
			
			\draw[line width=2pt,white] (inst3) -- ++(0,-9*\unitfactor-2.2*\unitfactor);
			
			\protosep{\textbf{Relay Attack}}
			\messbotharr{pcd}{Unlocking protocol messages}{adv}Capture message and record in log.
			\messnoarr{adv}{Capture, record, and forward message.}{adve}
			\messbotharr{adve}{Unlocking protocol messages}{picc}
			\stepcounter{seqlevel}
			
			\protosep{\textbf{Replay Attack}}
			\messbotharr{pcd}{Unlocking protocol messages}{adv}
			\messnoarr{adv}{Replay previously recorded messages.}{adve}
			\stepcounter{seqlevel}
			
			\protosep{\textbf{Walk-By Attack}}
			\mess{adve}{Get $UID$ }{picc}
			\stepcounter{seqlevel}
			\messnoarr{adv}{Save $UID$ from PICC.}{adve}
			\mess{picc}{$UID$ }{adve}
			\stepcounter{seqlevel}
			\messbotharr{pcd}{Unlocking protocol messages}{adv}
			\messnoarr{adv}{PICC role with $UID$.}{adve}
			\stepcounter{seqlevel}
			
			\protosep{\textbf{Privilege Escalation/Brute-Force Attack}}
			\messbotharr{pcd}{Unlocking protocol messages}{adv}
			\messnoarr{adv}{PICC role with guessed $UID$.}{adve}
		\end{sequencediagram}
		\caption{Attack setups with cylinder (PCD), transponder (PICC), and adversary $\adv$.}
		\label{proto:issues}
	\end{figure*}

	\subsection{Security Issues}
		In this section, we discuss security-related issues in the design and implementation of the cylinder unlocking procedure. Proof of concept attacks facilitate {\NFCG}'s features to demonstrate their applicability in real-world scenarios using off-the-shelf hardware. \autoref{proto:issues} shows the different attack setups. We consider the adversary $\adv$ in the Dolev-Yao model \autocite{dolevyao}, who can capture, block, modify, and resend captured messages all without breaking the cryptographic primitives.
		
		\paragraph{Relay Attack.} With {\NFCG}'s relay mode we are able to forward the \gls{NFC} traffic between \gls{PICC} and \gls{PCD} over a network. $\adv$ employs two smartphones running {\NFCG} connected to the server, one acting as the reader and the other as the tag. They communicate with the transponder and cylinder while transmitting \gls{NFC} traffic to the server. As long as the unlocking procedure takes less than $\approx 1.8\,\mathrm{s}$, it leads to a successful unlocking of the cylinder. Considering the average network delay around half the globe [Frankfurt/Main (Germany) to Sydney (Australia)] is $\approx 360\,\mathrm{ms}$ \autocite{ping} we could successfully unlock the cylinder from around the world. This demonstrates that the time limit imposed by the cylinder is not an effective countermeasure to these kinds of attacks. In addition to tightening the timings, \emph{distance bounding} techniques could be used as further mitigations (see \autoref{sec:db}).
		
		\paragraph{Replay Attack.} Using messages previously captured during an unlocking procedure between the cylinder and an authorized transponder, $\adv$ replays the traffic employing {\NFCG} in the tag role.
		As a result of the protocol modifications outlined in \autoref{proto:lock} and in contrast to the replay-protected DESFire authentication protocol, the unlocking procedure used by the cylinder is vulnerable to replaying the communication.
		Due to the static nonce $r_A$ used by the \gls{PCD} and the zero-initialized $IV_A$ and $IV_B$, the randomness of the protocol entirely depends on the nonce $r_B$ chosen by the \gls{PICC} as shown in \autoref{proto:lock}.
		
		Replaying \gls{PICC} messages to the \gls{PCD} results in the same randomness, leading to the same session key $k_s$. Since the encrypted channel relies only on $k_s$ and the content of the messages, which is always the same $UID$, an identical $k_s$ leads to the exact same messages. This replay attack does not require knowledge of any encryption key or breaking any primitive in the protocol. Conforming to the DESFire protocol by choosing the nonce $r_A$ randomly instead of statically mitigates replay attacks.
		
		\paragraph{Walk-By Attack.} Because the secret key $k$ used for encryption in the authentication phase is static for all cylinders and transponders, we can extract this key from the freely available utility software. This application allows the registration of new transponders to the system without requiring the physical presence of a cylinder by using a separately sold USB \gls{NFC} reader. The slightly obfuscated key $k$ can be extracted from the software by reverse-engineering the application binary. Therefore, we now consider $k$ to be known to $\adv$ and adapt our model to include the encryption and decryption of protocol messages.
		
		The static key allows the implementation of the \gls{PCD} role in the unlocking procedure as an {\NFCG} server plugin. Since the plugin reads the $UID$ of transponders without requiring access to a cylinder, it simplifies the relay attack discussed earlier. In this walk-by attack, $\adv$ can obtain the authorization by requesting the $UID$ from a transponder, without needing an active connection to the cylinder. Additionally, this allows the creation of multiple copies of a single transponder, thus violating the software design where every transponder is uniquely mapped to exactly one user.
		
		Walk-by attacks could be mitigated by using individual keys for each deployment, since the key is only available to users of this particular system. Even if an adversary $\adv$ had access to the key $k$ of their own system, no attacks on other systems would be possible. In addition, extracting $k$ from a transponder or cylinder is not trivial because it requires destructive physical access to the hardware.
		
		\paragraph{Privilege Escalation and Brute-Force Attack.} Comparing $UIDs$ obtained from different transponders in consecutive production batches indicates the presence of a pattern. The $UIDs$ have 7 bytes, where the first byte specifies the tag's manufacturer, in our case \texttt{04} for NXP. The remaining 6 bytes do not seem to be chosen at random, instead tags from the same batch have similar byte sequences. This suggests a serial number pattern. The numerical difference between the examined $UIDs$ of our two transponders with the same production date and consecutive batches is $3596$.
		
		If $\adv$ has access to a transponder in a system without authorization to unlock a specific cylinder, they can use the static key $k$ to extract the $UID$ of their transponder. This significantly reduces the number of guesses required for other $UIDs$ since tags of the system were likely produced in the same or close production batches. Guessing a $UID$ correctly escalates privileges of $\adv$ to other cylinders. A plugin for the {\NFCG} server implements brute-forcing $UIDs$, optionally starting from a known value. It achieves a throughput of approximately three tries per second since the cylinder implements no limit on the number of authorization tries. In our example, it would take around 20 minutes to execute such an attack.
		
		These attacks can be mitigated by using a random number as authorization instead of the predictable $UID$. This random number would be stored on the tag protected by the existing authentication protocol. Furthermore, using individual keys per base station instead of a static key $k$ hinders the execution of this attack. Adding brute-force countermeasures by limiting the number of authorization tries in a specific time frame significantly slows down any brute-force attempt.
		
	\subsection{Responsible Disclosure}
		We contacted the vendor to report our findings, and received an initial response within four hours. After providing the vendor with our report, we discussed the discovered issues in a telephone conference and an in-person meeting. The issues will be fixed in the next revision of the system, and the vendor aims to provide patches to existing customers. The entire process was handled in a professional and collaborative manner.

\section{Performance Evaluation}

	\glsreset{FWT}
	
	After demonstrating the capabilities of the software in this case study, we also evaluate the delays induced by {\NFCG} in relay and replay mode for more general cases.
	In this section, we investigate whether an upper bound on the response time imposed by the \gls{PCD} could be a viable protection mechanism against relay attacks by determining the additional latency in comparison to a direct communication between a \gls{PCD} and a \gls{PICC}.
	One such upper bound could be imposed by the \gls{FWT} as defined in ISO/IEC~14443.

	The evaluation consists of measuring the response time of a common command sequence for retrieving a value of a card in various \gls{PICC} configurations. We use a PN532 chipset connected to a computer via USB/UART as the \gls{PCD} and measure the command response time of the \gls{PICC}. The original Mifare DESFire tag operates in ISO/IEC~7816 \gls{ICS} mode. In particular, the measured commands are:
	\begin{enumerate}
		\item \gls{ICS} \texttt{SELECT} file: DESFire AID (\texttt{0xA4})
		\item Select Application (\texttt{0x5A})
		\item Get FileSettings (\texttt{0xF5})
		\item Get Value (\texttt{0x6C})
	\end{enumerate}
	
	The {\NFCG} app runs on one Nexus 5X (Android 7) smartphone in the tag role and one OnePlus 6 (Android 9) smartphone in the reader role, while the {\NFCG} server is hosted directly on the Nexus 5X or a computer. In addition to a baseline measurement, where the \gls{PCD} communicates with the original tag (TAG), we measure the latency in the following configurations:
	\begin{enumerate}
		\item[RP] \emph{Local replay}. One smartphone replays previously captured \gls{NFC} traffic locally to the \gls{PCD}. No additional smartphone or server is used.
		\item[BT] \emph{Bluetooth relay}. Smartphones connected via a Bluetooth PAN. The server is hosted on one of the smartphones.
		\item[BW] \emph{Bluetooth tethering to wireless network}. One smartphone provides IEEE~802.11 wireless network access to the other via Bluetooth tethering. The server is hosted on a computer wired to the network.
		\item[WH] \emph{IEEE~802.11 wireless hotspot}. One smartphone offers a wireless network access point to the other one. The server is hosted on the smartphone providing the access point.
		\item[WA] \emph{IEEE~802.11 wireless network}. Both smartphones are connected to the same wireless network. The server is hosted on a computer wired to the network.
	\end{enumerate}

	Every configuration using wireless technologies is conducted in an urban environment with multiple other wireless networks in active use.

	\begin{figure}
		\resizebox{\linewidth}{!}{\input{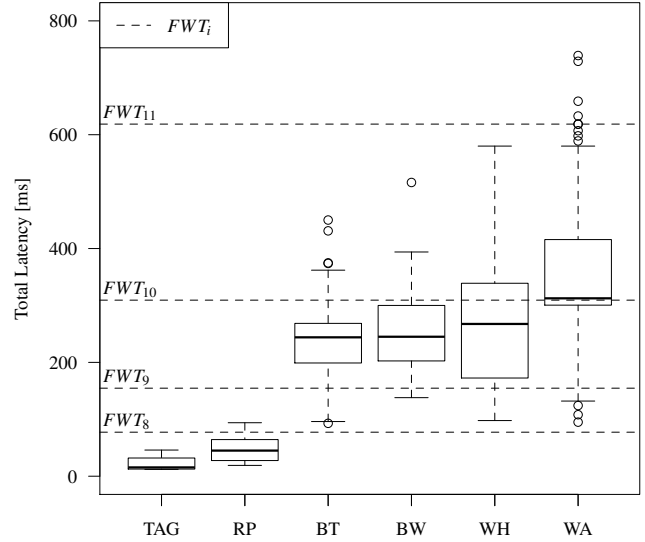}}
		\caption{Latency measurements using NFCGate.}
		\label{fig:eval}
	\end{figure}

	\autoref{fig:eval} depicts the total latency of command responses for the different configurations as a box plot ($n = 20$) with outliers, minima, maxima, median, and the lowest/highest values within $1.5$ interquartile range. Dashed lines show values for the $\gls{FWT}$, where $FWT_j$ is only drawn for $8 \le j \le 11$, because $FWT_8$ is the minimum $FWT$ specified by the original tag and any $FWT_j$ with $j \ge 12$ is out of bounds. For $FWT \ge FWT_{11} \approx 619\,\mathrm{ms}$, the latency stays within $\gls{FWT}$ bounds for all tested configurations, excluding some outliers. We note that the replay mostly stays within the $\gls{FWT}$ of the original tag and is in some cases indistinguishable from it.
	
	All network configurations using the IEEE~802.11 wireless network standard with both smartphones exhibit a high variance in latency measurements while Bluetooth-based configurations appear to be more stable. We attribute the high variance to interference in the IEEE~802.11 wireless network and the stable Bluetooth measurements to the low data throughput in our evaluation. Therefore, we recommend a Bluetooth PAN configuration for close proximity relays.
	
\section{Countermeasures}\label{sec:db}
	The $\gls{FWT}$ has been proposed as a potential countermeasure to relay attacks \autocite{rl:jose:payment,rl:francis:payment}, although it was primarily designed as a safety measure. Despite its standardization, the $\gls{FWT}$ has not been enforced in our experiments, which seems to be very common \autocite{rl:bocek:payment}. Even if it was enforced, the $\gls{FWT}$ could be increased by the \gls{PICC} up to $\approx 5\,\mathrm{s}$. In the event of a loose mandatory $\gls{FWT}$ enforced by the \gls{PCD}, our measurements show that it is possible to relay the communication even if $FWT_j$ for $j \ge 10$ is in use. Assuming the use of the original tag's $\gls{FWT}$ is mandatory, a replay could be performed nevertheless. Especially in the case of a cryptographically intensive operation, a replay could be indistinguishable in timing from the original tag, since it does not use cryptography. Despite some configurations exceeding the $\gls{FWT}$, a relay could still be performed, because the $\gls{FWT}$ was designed as a safety measure, not a security feature. The \gls{PCD} simply retransmits a block after the $\gls{FWT}$ expires, allowing even more time for a relay to respond \autocite[Section~7.2]{std:iso:14443}. This shows that $\gls{FWT}$ is not just an ineffective countermeasure but no countermeasure at all.

	\paragraph{Mandatory Response Timeouts.}
	One possible countermeasure against relay attacks is the implementation of a tight mandatory response timeout on the reader side. In contrast to the \gls{FWT}, this cannot be influenced by the tag and constitutes a definitive communication timeout.
	Since this countermeasure requires knowing an upper limit on the response latency to any connectable tag beforehand, 
	it only works in closed ecosystems, where all involved readers and tags are controllable. The duration of cryptographic operations in particular depends on hardware instruction support and performance, so low-end \gls{HCE} devices might raise the upper limit on response latency, which would in turn reduce the effectiveness of this countermeasure.
	
	\paragraph{Distance Bounding Protocols.}
	First introduced by Brands and Chaum \autocite{rl:brands:db}, \gls{DB} protocols solve the general issue of relay attacks. In a \gls{DB} protocol, physical properties of a connection are used to prove that one endpoint is within a specified distance of the other. In addition to timing and freshness, some \gls{DB} protocols employ cryptography to ensure the authenticity of endpoints \autocite{rl:boureanu:db2, rl:boureanu:db}.
	
	\gls{DB} for \gls{NFC} can be implemented on either the protocol or application layer.
	An implementation on the protocol layer requires an extension to the ISO/IEC~14443 standard and performs \gls{DB} independently of the application \autocite{rl:drimer:db,rl:hancke:db,rl:reid:db}. While this simplifies the usage for many applications, it only measures the distance to some recipient, not necessarily the intended one.
	In contrast, an implementation on the application layer can ensure the correct recipient, but requires every application to implement their own, incompatible \gls{DB} protocol \autocite{rl:hermans:db,rl:chothia:db,rl:kilinc:db}.
	
	One example for an application layer \gls{DB} protocol is the DESFire EV2 Proximity Check \autocite{rl:smith:ev2}. The check uses cryptography to ensure mutual authentication at the end of the timing phase. Despite these engineering efforts, an attack against the \gls{DB} protocol has already been discovered \autocite{rl:celiano:ev2}.
	
\section{Conclusion}
	With {\NFCG}, we have enhanced the original \gls{PoC} to an extensive \gls{NFC} research toolkit that is compatible with any Android smartphones supporting the Xposed or EdXposed hooking framework and requires no changes to the system image. The interoperability proved invaluable in the case study, since it allowed us to further inspect the captured traffic in Wireshark with its existing \gls{NFC} protocol dissectors. Using the toolkit, we analyzed a well-known \gls{NFC}-based door locking system and uncovered several security issues of various severities. We launched a brute-force attack against the $UID$ used as authorization employing the {\NFCG} server plugin feature. By informing the vendor of our case study, several security issues could be fixed, which helped secure a currently deployed product.
	The evaluation showed that the relay latency is low enough to bypass many countermeasures, while the negligible replay latency makes it almost indistinguishable from the original tag.

	As future work, the Wireshark remote capturing protocol could be implemented, streaming the captured data from {\NFCG} to a running live capture to simplify the use of the \emph{pcapng} export.
	Finally, we suggest using the on-device capture mode to further analyze \gls{HCE}-based apps such as Google Pay, or inspect \gls{NFC} reader apps, e.g. public transport services.
	
\section*{Availability}
We make the source code of {\NFCG} available to the public under a free software license \autocite{nfcgate:code}.

\section*{Acknowledgments}
This work has been co-funded by the DFG as part of project C.1 within the RTG 2050 ``Privacy and Trust for Mobile Users'', and by the German Federal Ministry of Education and Research and the Hessen State Ministry for Higher Education, Research and the Arts within their joint support of the National Research Center for Applied Cybersecurity ATHENE, and by the LOEWE initiative (Hesse, Germany) within the emergenCITY centre.

	{\sloppy
	\printbibliography}

@techreport{spec:pcapng,
	author = {Tuexen, Michael and Risso, Fulvio and Bongertz, Jasper and Combs, Gerald and Harris, Guy},
	date = {2020-01-26},
	howpublished = {Internet Requests for Comments},
	institution = {Network Working Group},
	journaltitle = {Internet-Draft},
	publisher = {Network Working Group},
	subtitle = {draft-tuexen-opsawg-pcapng},
	title = {{PCAP} {Next} {Generation} (pcapng) capture file format},
	type = {Internet-Draft}
}

@inproceedings{rl:boureanu:db2,
	address = {Cham},
	author = {Boureanu, Ioana and Mitrokotsa, Aikaterini and Vaudenay, Serge},
	booktitle = {Information Security},
	editor = {Desmedt, Yvo},
	isbn = {978-3-319-27659-5},
	pages = {248–258},
	publisher = {Springer International Publishing},
	title = {Practical and Provably Secure Distance-Bounding},
	year = {2015}
}

@inproceedings{rl:brands:db,
	author = {Brands, Stefan and Chaum, David},
	booktitle = {Workshop on the Theory and Application of of Cryptographic Techniques},
	organization = {Springer},
	pages = {344–359},
	title = {Distance-bounding protocols},
	year = {1993}
}

@inproceedings{rl:francis:payment,
	author = {Francis, Lishoy and Hancke, Gerhard and Mayes, Keith and Markantonakis, Konstantinos},
	booktitle = {Radio Frequency Identification System Security},
	doi = {10.3233/978-1-61499-143-4-21},
	isbn = {9781614991427},
	language = {English},
	pages = {21–32},
	series = {Cryptology and Information Security Series},
	title = {Practical Relay Attack on Contactless Transactions by Using {NFC} Mobile Phones},
	volume = {8},
	year = {2012}
}

@techreport{std:iso:14443-1,
	address = {Geneva, CH},
	author = {{ISO/IEC 14443-1:2000}},
	edition = {1},
	institution = {International Organization for Standardization},
	month = 04,
	pagetotal = {5},
	title = {Identification cards – Contactless integrated circuit(s) cards – Proximity cards – Part 1: Physical characteristics},
	type = {Standard},
	year = {2000}
}

@techreport{std:iso:7816-3,
	address = {Geneva, CH},
	author = {{ISO/IEC 7816-3:2006}},
	edition = {3},
	institution = {International Organization for Standardization},
	month = 11,
	pagetotal = {58},
	title = {Identification cards – Integrated circuit cards – Part 3: Cards with contacts – Electrical interface and transmission protocols},
	type = {Standard},
	year = {2006}
}

@techreport{std:iso:7816-4,
	address = {Geneva, CH},
	author = {{ISO/IEC 7816-4:2005}},
	edition = {2},
	institution = {International Organization for Standardization},
	month = 01,
	pagetotal = {90},
	title = {Identification cards – Integrated circuit cards – Part 4: Organization, security and commands for interchange},
	type = {Standard},
	year = {2005}
}

@online{ws:iso14443,
	author = {Kaiser, Martin and Harris, Guy and Combs, Gerald and others},
	subtitle = {wireshark/wireshark},
	title = {wireshark/packet-iso14443.c at wireshark-3.2.1},
	url = {https://github.com/wireshark/wireshark/blob/wireshark-3.2.1/epan/dissectors/packet-iso14443.c},
	urldate = {2020-01-27},
	year = {2019}
}

@online{rl:android:hce,
	author = {{Android Open Source Project}},
	date = {2019-12-27},
	subtitle = {Android Developers},
	title = {Host-based card emulation overview},
	url = {https://developer.android.com/guide/topics/connectivity/nfc/hce},
	urldate = {2020-02-24}
}

@online{ws:iso7816,
	author = {Kaiser, Martin and Harris, Guy and Combs, Gerald and others},
	subtitle = {wireshark/wireshark},
	title = {wireshark/packet-iso7816.c at wireshark-3.2.1},
	url = {https://github.com/wireshark/wireshark/blob/wireshark-3.2.1/epan/dissectors/packet-iso7816.c},
	urldate = {2020-01-27},
	year = {2019}
}

@article{dolevyao,
	author = {{Dolev}, Danny and {Yao}, Andrew C.},
	doi = {10.1109/TIT.1983.1056650},
	issn = {1557-9654},
	journal = {IEEE Transactions on Information Theory},
	month = {03},
	number = {2},
	pages = {198–208},
	title = {On the security of public key protocols},
	volume = {29},
	year = {1983}
}

@online{ping,
	author = {{WonderNetwork}},
	subtitle = {WonderNetwork},
	title = {Ping time between {Frankfurt} and other cities},
	url = {https://wondernetwork.com/pings/Frankfurt},
	urldate = {2020-02-06},
	year = {2020}
}

@online{nfcgate:code,
	author = {{The NFCGate Team}},
	subtitle = {An {NFC} research toolkit application for {Android}},
	title = {{NFCGate}},
	url = {https://github.com/nfcgate/nfcgate},
	urldate = {2020-03-19},
	year = {2020}
}

@online{chameleon,
	author = {{Kasper \& Oswald GmbH}},
	title = {{ChameleonMini}},
	url = {https://kasper-oswald.de/gb/chameleonmini/},
	urldate = {2020-02-13},
	year = {2020}
}

@online{code:native,
	author = {{Android Open Source Project}},
	date = {2020-04-16},
	title = {nfc - Source},
	url = {https://cs.android.com/android/_/android/platform/system/nfc/+/91688f6:src/nfc/},
	urldate = {2020-05-18}
}

@online{code:transceive,
	author = {{Android Open Source Project}},
	date = {2020-04-09},
	title = {{NfcService.java} - Source},
	url = {https://cs.android.com/android/_/android/platform/packages/apps/Nfc/+/b9d4425:src/com/android/nfc/NfcService.java;l=1470},
	urldate = {2020-05-18}
}

@online{code:dispatch,
	author = {{Android Open Source Project}},
	date = {2020-03-24},
	title = {{NfcDispatcher.java} - Source},
	url = {https://cs.android.com/android/_/android/platform/packages/apps/Nfc/+/68fe178:src/com/android/nfc/NfcDispatcher.java;l=270},
	urldate = {2020-05-18}
}

@online{code:hce,
	author = {{Android Open Source Project}},
	date = {2019-05-15},
	title = {{HostEmulationManager.java} - Source},
	url = {https://cs.android.com/android/_/android/platform/packages/apps/Nfc/+/ac65163:src/com/android/nfc/cardemulation/HostEmulationManager.java},
	urldate = {2020-05-18}
}

@online{proxmark,
	author = {{ProxGrind}},
	title = {{Proxmark3} {Rdv4.0} Development},
	url = {https://proxgrind.com/prototyping/proxmark3-rdv4-0-development/},
	urldate = {2020-02-13},
	year = {2020}
}

@inproceedings{rl:boureanu:db,
	author = {Boureanu, Ioana and Mitrokotsa, Aikaterini and Vaudenay, Serge},
	booktitle = {International Workshop on Fast Software Encryption},
	organization = {Springer},
	pages = {55–67},
	title = {Towards Secure Distance Bounding},
	year = {2013}
}

@online{rl:qz:apay,
	author = {{Quartz}},
	date = {2020-02-11},
	title = {{Apple} {Pay} is on pace to account for 10\% of all global card transactions},
	url = {https://qz.com/1799912/apple-pay-on-pace-to-account-for-10-percent-of-global-card-transactions/},
	urldate = {2020-02-12}
}

@online{rl:pay:market,
	author = {{yStats GmbH \& Co. KG}},
	date = {2019-03-18},
	pagetotal = {45},
	title = {Mobile Wallet Profiles 2019: {Apple} {Pay}, {Google} {Pay}, {Samsung} {Pay}},
	url = {https://www.ystats.com/market-reports/google-pay-profile-2019},
	urldate = {2020-02-12}
}

@online{rl:apple:nfc,
	author = {{Apple Inc.}},
	subtitle = {Apple Developer Documentation},
	title = {Core {NFC}},
	url = {https://developer.apple.com/documentation/corenfc},
	urldate = {2020-02-12},
	year = {2019}
}

@online{rl:android:nfc,
	author = {{Android Open Source Project}},
	date = {2019-12-27},
	subtitle = {Android Developers},
	title = {Near field communication overview},
	url = {https://developer.android.com/guide/topics/connectivity/nfc},
	urldate = {2020-02-12}
}

@misc{rl:giese:payment,
	archiveprefix = {arXiv},
	author = {Giese, Dennis and Liu, Kevin and Sun, Michael and Syed, Tahin and Zhang, Linda},
	eprint = {1904.10623},
	primaryclass = {cs.CR},
	title = {Security Analysis of {Near-Field} {Communication (NFC)} Payments},
	year = {2019}
}

@inproceedings{rl:jose:payment,
	address = {Cham},
	author = {Vila, José and Rodríguez, Ricardo J.},
	booktitle = {Radio Frequency Identification},
	editor = {Mangard, Stefan and Schaumont, Patrick},
	isbn = {978-3-319-24837-0},
	pages = {87–103},
	publisher = {Springer International Publishing},
	title = {Practical Experiences on {NFC} Relay Attacks with {Android}},
	year = {2015}
}

@inproceedings{rl:bocek:payment,
	address = {Cham},
	author = {Bocek, Thomas and Killer, Christian and Tsiaras, Christos and Stiller, Burkhard},
	booktitle = {Management and Security in the Age of Hyperconnectivity},
	editor = {Badonnel, Rémi and Koch, Robert and Pras, Aiko and Drašar, Martin and Stiller, Burkhard},
	isbn = {978-3-319-39814-3},
	pages = {71–83},
	publisher = {Springer International Publishing},
	title = {An {NFC} Relay Attack with Off-the-shelf Hardware and Software},
	year = {2016}
}

@inproceedings{rl:roland:payment,
	author = {{Roland}, Michael and {Langer}, Josef and {Scharinger}, Josef},
	booktitle = {2013 5th International Workshop on Near Field Communication (NFC)},
	doi = {10.1109/NFC.2013.6482441},
	issn = {null},
	month = {02},
	pages = {1–6},
	title = {Applying relay attacks to {Google} {Wallet}},
	year = {2013}
}

@inproceedings{rl:roland:payment2,
	address = {Washington, D.C.},
	author = {Roland, Michael and Langer, Josef},
	booktitle = {Presented as part of the 7th {USENIX} Workshop on Offensive Technologies},
	publisher = {{USENIX}},
	title = {Cloning Credit Cards: A Combined Pre-play and Downgrade Attack on {EMV} {Contactless}},
	url = {https://www.usenix.org/conference/woot13/workshop-program/presentation/Roland},
	year = {2013}
}

@misc{rl:galloway:payment,
	author = {Galloway, Leigh-Anne and Yunusov, Tim and Stennikov, Aleksei},
	date = {2019-12-04},
	title = {First Contact: New Vulnerabilities in Contactless Payments},
	url = {https://leigh-annegalloway.com/presentation-materials/},
	urldate = {2020-02-13}
}

@online{sw:wireshark,
	author = {Combs, Gerald and Ramirez, Gilbert and Harris, Guy and others},
	title = {Wireshark},
	url = {https://www.wireshark.org/},
	urldate = {2020-02-13},
	year = {2020}
}

@online{code:nfctech,
	author = {{Android Open Source Project}},
	date = {2020-01-22},
	title = {tech - Source},
	url = {https://cs.android.com/android/_/android/platform/frameworks/base/+/3eb265a:core/java/android/nfc/tech},
	urldate = {2020-02-20}
}

@online{code:hal,
	author = {{Android Open Source Project}},
	date = {2020-02-10},
	title = {halimpl - Source},
	url = {https://cs.android.com/android/_/android/platform/hardware/nxp/nfc/+/65c90b3:halimpl},
	urldate = {2020-02-20}
}

@online{rl:google:pay,
	author = {{Google LLC}},
	subtitle = {{Google Pay} Help},
	title = {Set up {Google Pay} – {Android}},
	url = {https://support.google.com/pay/answer/7625055?co=GENIE.Platform%3DAndroid},
	urldate = {2020-02-13},
	year = {2020}
}

@inproceedings{nfcproxy,
	author = {Lee, Eddie},
	booktitle = {Defcon hacking conference},
	pages = {63–74},
	title = {{NFC} hacking: The easy way},
	volume = {20},
	year = {2012}
}

@misc{dalheimer:eauto,
	author = {Dalheimer, Mathias},
	date = {2017-12-27},
	eventtitle = {34C3},
	institution = {Chaos Computer Club e.V},
	title = {{Warum das Laden eines Elektroautos unsicher ist}},
	url = {https://media.ccc.de/v/34c3-9092-ladeinfrastruktur_fur_elektroautos_ausbau_statt_sicherheit},
	urldate = {2020-02-17}
}

@misc{rl:yuyi:range,
	archiveprefix = {arXiv},
	author = {Sun, Yuyi and Kumar, Swarun and He, Shibo and Chen, Jiming and Shi, Zhiguo},
	eprint = {2001.08143},
	primaryclass = {cs.CR},
	title = {You foot the bill! {Attacking} {NFC} with passive relays},
	year = {2020}
}

@techreport{std:iso:14443,
	address = {Geneva, CH},
	author = {{ISO/IEC 14443-4:2001}},
	edition = {1},
	institution = {International Organization for Standardization},
	month = 02,
	pagetotal = {34},
	title = {Identification cards – Contactless integrated circuit(s) cards – Proximity cards – Part 4: Transmission protocol},
	type = {Standard},
	year = {2001}
}

@techreport{std:jis:6319,
	address = {Tokyo, JP},
	author = {{JSA - JIS X 6319-4}},
	institution = {Japanese Standards Association},
	month = 03,
	pagetotal = {129},
	title = {Specification of implementation for integrated circuit(s) cards – Part 4: High speed proximity cards},
	type = {Standard},
	year = {2016}
}

@techreport{std:iso:15693,
	address = {Geneva, CH},
	author = {{ISO/IEC 15693-1:2010}},
	edition = {2},
	institution = {International Organization for Standardization},
	month = 10,
	pagetotal = {5},
	title = {Identification cards – Contactless integrated circuit cards – Vicinity cards – Part 1: Physical characteristics},
	type = {Standard},
	year = {2010}
}

@techreport{std:desfire,
	author = {{NXP Semiconductors N.V.}},
	date = {2005-12-09},
	edition = {3.2},
	pagetotal = {18},
	subtitle = {MIFARE DESFire EV1 contactless multi-application IC},
	title = {MF3ICDx21\_41\_81},
	type = {Product short data sheet},
	url = {https://www.nxp.com/docs/en/data-sheet/MF3ICDX21_41_81_SDS.pdf},
	urldate = {2020-02-23}
}

@techreport{std:nci,
	author = {{NFC Forum}},
	edition = {2.0},
	month = {11},
	pagetotal = {206},
	title = {{NFC Controller Interface (NCI)}},
	type = {Technical Specification},
	year = {2016}
}

@online{sw:xposed,
	author = {{rovo89} and {Thungstwenty}},
	subtitle = {{Xposed} Module Repository},
	title = {Welcome to the {Xposed} Module Repository},
	url = {https://xposed.info},
	urldate = {2020-02-24},
	year = {2020}
}

@online{sw:edxposed,
	author = {{solohsu} and {Jim Wu}},
	subtitle = {{MeowCat Studio}},
	title = {{EdXposed} Framework Official Website},
	url = {https://edxp.meowcat.org/},
	urldate = {2020-02-24},
	year = {2020}
}

@inproceedings{rl:drimer:db,
	address = {USA},
	articleno = {Article 7},
	author = {Drimer, Saar and Murdoch, Steven J.},
	booktitle = {Proceedings of 16th USENIX Security Symposium on USENIX Security Symposium},
	isbn = {1113335555779},
	location = {Boston, MA},
	numpages = {16},
	publisher = {USENIX Association},
	series = {SS’07},
	title = {Keep Your Enemies Close: Distance Bounding against Smartcard Relay Attacks},
	year = {2007}
}

@inproceedings{rl:hancke:db,
	address = {USA},
	author = {Hancke, Gerhard P. and Kuhn, Markus G.},
	booktitle = {Proceedings of the First International Conference on Security and Privacy for Emerging Areas in Communications Networks},
	doi = {10.1109/SECURECOMM.2005.56},
	isbn = {0769523692},
	numpages = {7},
	pages = {67–73},
	publisher = {IEEE Computer Society},
	series = {SECURECOMM ’05},
	title = {An {RFID} Distance Bounding Protocol},
	url = {https://doi.org/10.1109/SECURECOMM.2005.56},
	year = {2005}
}

@inproceedings{rl:reid:db,
	address = {New York, NY, USA},
	author = {Reid, Jason and Nieto, Juan M. Gonzalez and Tang, Tee and Senadji, Bouchra},
	booktitle = {Proceedings of the 2nd ACM Symposium on Information, Computer and Communications Security},
	doi = {10.1145/1229285.1229314},
	isbn = {1595935746},
	location = {Singapore},
	numpages = {10},
	pages = {204–213},
	publisher = {Association for Computing Machinery},
	series = {ASIACCS ’07},
	title = {Detecting Relay Attacks with Timing-Based Protocols},
	url = {https://doi.org/10.1145/1229285.1229314},
	year = {2007}
}

@inproceedings{rl:hermans:db,
	address = {New York, NY, USA},
	author = {Hermans, Jens and Peeters, Roel and Onete, Cristina},
	booktitle = {Proceedings of the Sixth ACM Conference on Security and Privacy in Wireless and Mobile Networks},
	doi = {10.1145/2462096.2462129},
	isbn = {9781450319980},
	location = {Budapest, Hungary},
	numpages = {12},
	pages = {207–218},
	publisher = {Association for Computing Machinery},
	series = {WiSec ’13},
	title = {Efficient, Secure, Private Distance Bounding without Key Updates},
	url = {https://doi.org/10.1145/2462096.2462129},
	year = {2013}
}

@inproceedings{rl:chothia:db,
	address = {Berlin, Heidelberg},
	author = {Chothia, Tom and Garcia, Flavio D. and de Ruiter, Joeri and van den Breekel, Jordi and Thompson, Matthew},
	booktitle = {Financial Cryptography and Data Security},
	editor = {Böhme, Rainer and Okamoto, Tatsuaki},
	isbn = {978-3-662-47854-7},
	pages = {189–206},
	publisher = {Springer Berlin Heidelberg},
	title = {Relay Cost Bounding for Contactless {EMV} Payments},
	year = {2015}
}

@inproceedings{rl:kilinc:db,
	address = {Cham},
	author = {Kılınç, Handan and Vaudenay, Serge},
	booktitle = {Information Security},
	editor = {Nguyen, Phong Q. and Zhou, Jianying},
	isbn = {978-3-319-69659-1},
	pages = {195–213},
	publisher = {Springer International Publishing},
	title = {Contactless Access Control Based on Distance Bounding},
	year = {2017}
}

@inproceedings{nfcgate,
	address = {New York, NY, USA},
	articleno = {Article 27},
	author = {Maass, Max and Müller, Uwe and Schons, Tom and Wegemer, Daniel and Schulz, Matthias},
	booktitle = {Proceedings of the 8th ACM Conference on Security \& Privacy in Wireless and Mobile Networks},
	doi = {10.1145/2766498.2774984},
	isbn = {9781450336239},
	location = {New York, New York},
	numpages = {2},
	publisher = {Association for Computing Machinery},
	series = {WiSec ’15},
	title = {{DEMO}: {NFCGate}: An {NFC} Relay Application for {Android}},
	url = {https://doi.org/10.1145/2766498.2774984},
	year = {2015}
}

@online{sw:freefare,
	author = {Conty, Romuald and Tartière, Romain and Teuwen, Philippe and others},
	subtitle = {A convenience {API} for {NFC} cards manipulations on top of libnfc},
	title = {nfc-tools/libfreefare},
	url = {https://github.com/nfc-tools/libfreefare/},
	urldate = {2020-03-06}
}

@inproceedings{rl:smith:ev2,
	author = {Hurley-Smith, Darren and Hernandez-Castro, Julio},
	booktitle = {Cryptacus 2017},
	title = {Measuring the Distance: Investigating the {DESFire} {EV2} Distance Bounding Protocol},
	year = {2017}
}

@mastersthesis{rl:celiano:ev2,
	author = {Celiano, Dominic},
	institution = {University of Cambridge},
	title = {Overclocking Proximity Checks in Contactless Smartcards},
	year = {2018}
}

@article{rl:lopez:hooking,
	author = {Lopez, Juan and Babun, Leonardo and Aksu, Hidayet and Uluagac, A Selcuk},
	doi = {10.1007/s41635-017-0013-2},
	journal = {Journal of Hardware and Systems Security},
	number = {2},
	pages = {114–136},
	publisher = {Springer},
	title = {A survey on function and system call hooking approaches},
	volume = {1},
	year = {2017}
}

@inproceedings{rl:vogel:hooking,
	address = {San Diego, CA},
	author = {Vogl, Sebastian and Gawlik, Robert and Garmany, Behrad and Kittel, Thomas and Pfoh, Jonas and Eckert, Claudia and Holz, Thorsten},
	booktitle = {23rd {USENIX} Security Symposium ({USENIX} Security 14)},
	isbn = {978-1-931971-15-7},
	month = aug,
	pages = {813–328},
	publisher = {{USENIX} Association},
	title = {Dynamic Hooks: Hiding Control Flow Changes within Non-Control Data},
	url = {https://www.usenix.org/conference/usenixsecurity14/technical-sessions/presentation/vogl},
	year = {2014}
}

@article{rl:kim:hooking,
	author = {Kim, Sungkwan and Park, Junyoung and Lee, Kyungroul and You, Ilsun and Yim, Kangbin},
	journal = {J. Internet Serv. Inf. Secur.},
	number = {3/4},
	pages = {134–147},
	title = {A Brief Survey on Rootkit Techniques in Malicious Codes},
	volume = {2},
	year = {2012}
}

@online{sw:xhook,
	author = {{iQIYI Inc.}},
	subtitle = {A {PLT} hook library for {Android} native {ELF}},
	title = {{xHook}},
	url = {https://github.com/iqiyi/xHook},
	urldate = {2020-05-24}
}
\end{document}